\documentclass{IOS-Book-Article}
\usepackage[T1]{fontenc}
\usepackage[utf8]{inputenc}
\usepackage{graphicx}
\usepackage{mathptmx}
\usepackage{paralist}
\usepackage{epstopdf}
\usepackage{paralist}
\usepackage{tikz}
\usepackage{ctable}
\usepackage{caption,subcaption}
\usepackage{csquotes}
\usepackage{mathtools}
\usepackage[numbers]{natbib}
\usepackage[textsize=tiny]{todonotes}
\usepackage{url}
\def\hb{\hbox to 10.7 cm{}}

\newcommand{\dara}{\textsf{da\textbar ra}}
\begin{document}

\pagestyle{headings}
\def\thepage{}

\begin{frontmatter}
\title{A Semi-Automatic Approach for Detecting Dataset References in Social Science Texts}
\author[A,B]{\fnms{Behnam} \snm{Ghavimi}
\thanks{Corresponding Author: E-mail: behnam.ghavimi@gesis.org}},
\author[A]{\fnms{Philipp} \snm{Mayr}
\thanks{E-mail: philipp.mayr@gesis.org}},
\author[B,C]{\fnms{Christoph} \snm{Lange} 
\thanks{E-mail: langec@cs.uni-bonn.de}},
\author[B]{\fnms{Sahar} \snm{Vahdati}
\thanks{E-mail: vahdati@cs.uni-bonn.de}}
and
\author[B,C]{\fnms{S\"oren} \snm{Auer} 
\thanks{E-mail: auer@cs.uni-bonn.de}}

\address[A]{GESIS – Leibniz Institute for the Social Sciences}
\address[B]{Enterprise Information Systems (EIS), University of Bonn}
\address[C]{Fraunhofer Institute for Intelligent Analysis and Information Systems IAIS}
\begin{abstract}
Today, full-texts of scientific articles are often stored in different locations than the used datasets.
Dataset registries aim at a closer integration by making datasets citable 
but authors typically refer to datasets using inconsistent abbreviations and heterogeneous metadata (e.g. title, publication year).
It is thus hard to reproduce research results, to access datasets for further analysis, and to determine the impact of a dataset.
Manually detecting references to datasets in scientific articles is time-consuming and requires expert knowledge in the underlying research domain. 
We propose and evaluate a semi-automatic three-step approach for finding explicit references to datasets in social sciences articles.
We first extract pre-defined special features from dataset titles in the {\dara} registry, then detect references to datasets using the extracted features, and finally match the references found with corresponding dataset titles.
The approach does not require a corpus of articles (avoiding the cold start problem) and performs well on a test corpus. 
We achieved an F-measure of 0.84 for detecting references in full-texts and an F-measure of 0.83 for finding correct matches of detected references in the {\dara} dataset registry.

\end{abstract}

\begin{keyword}
Information extraction\sep Link discovery\sep Data linking\sep Research data\sep Social Sciences\sep Scientific articles\sep Data registry
\end{keyword}
\end{frontmatter}
\section{Introduction}

By its very nature science is based on evidence and facts.
In many scenarios, these emerge from data obtained from measurements, surveys, experiments or simulations.
As a result, scientific articles -- the key artifact of scholarly communication -- must refer to the ground truth the findings are based on, i.e. data.
Making these data references explicit, machine-readable and semantics-aware is one of the major challenges in digitizing scholarly communication effectively.
Especially in the quantitative social sciences and also most other fields of science many articles reference research datasets which they are based on.
For example, an article might present the results of a statistical analysis performed on a dataset comprising employment data in the European Union.
Today, digital libraries aim at providing resources with high-quality metadata, easy subject access, and further support and guidance for retrieving information~\citep{Hienert2015}. 
However, dataset references are usually not explicitly exposed in digital libraries.
In most cases the articles do not provide explicit links that give readers direct access to the referenced datasets. 

Explicit links from scientific publications to the underlying datasets and vice versa are useful in many scenarios, including:
\begin{itemize}
	\item reviewers aiming to reproduce the evaluation that authors performed on a dataset, 
	\item other researchers desiring to perform further analysis on a dataset that was used in an article,
	\item decision makers seeking to determine the impact of a given dataset or to identify the most used datasets in a given community.
\end{itemize}

Currently, the majority of published articles lack such direct links to datasets.
While there exist registries that make datasets citable, e.g., by assigning them a digital object identifier (DOI), they are usually not integrated with authoring tools.
Therefore, in practice, authors typically cite datasets by \emph{mentioning} them, e.g., using  
combinations of title, abbreviation and year of publication (see, e.g., \citet{Mathiak2015}).  

\begin{figure}[h]
	\centering
	\includegraphics[width=3.5in]{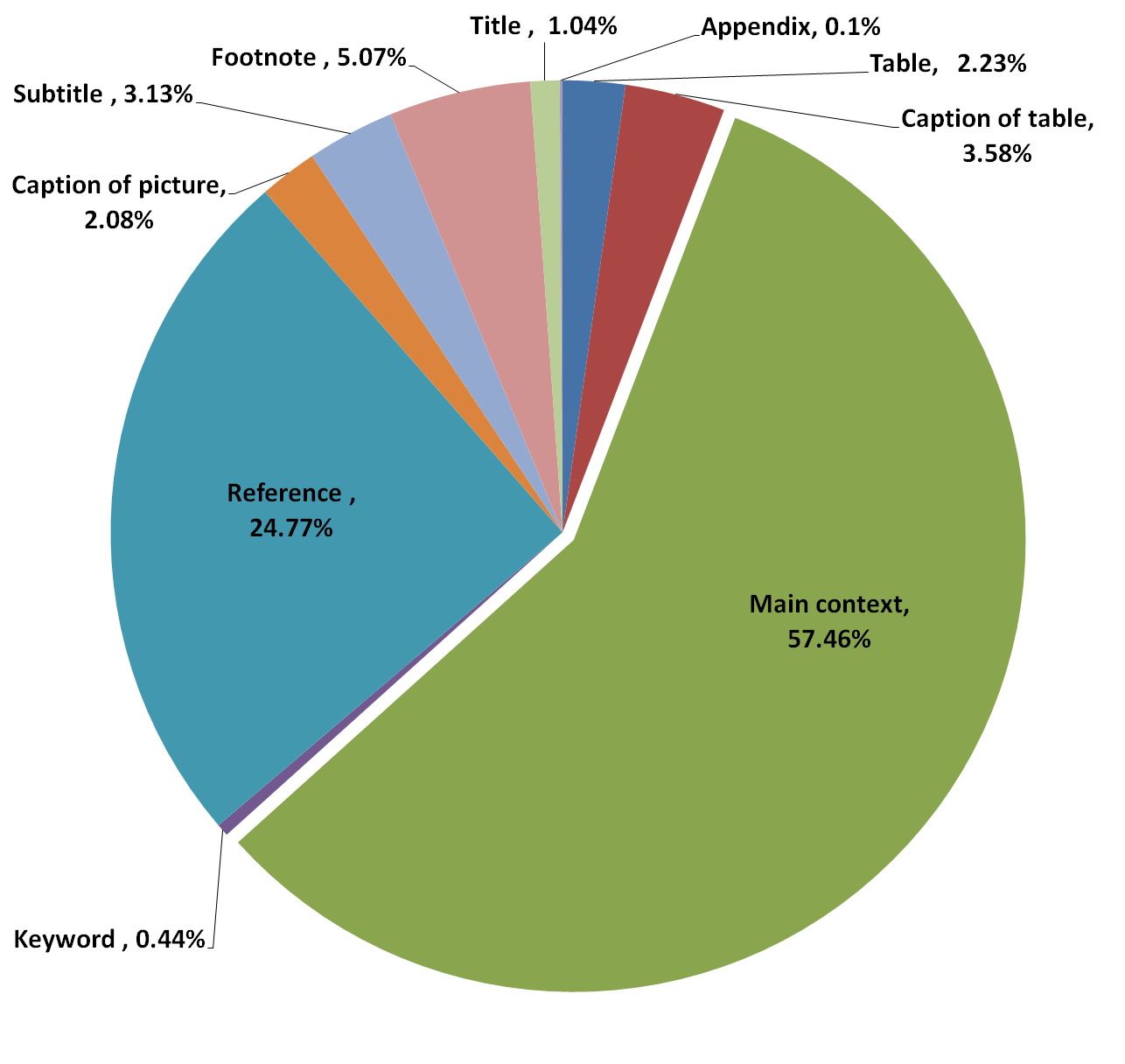}
	\caption{The distribution of more than 500 dataset references in 15 random articles from the mda journal (see Section~\ref{sec:mda})}
	\label{fig:places-example}
\end{figure}

Manually detecting references to datasets in articles is time consuming and requires expert knowledge of the article's domain. 
Detecting dataset references automatically is challenging 
due to the wide variety of styles of dataset citations in full-texts even within one research community, and the variety
of places in which datasets can be referenced in articles  
(illustrated in Figure~\ref{fig:places-example}).  
So it is a difficult task to create a training set manually for solving this issue and this variance even makes rule-based approaches difficult, as it is hard to cover all cases.
We therefore introduce a semi-automatic approach that parses full-texts and finds exact matches with high precision and without requiring a training set. 

The remainder of this section states the problem addressed by our research more precisely.
Section~\ref{sec:preliminaries} introduces the preliminaries of techniques and metrics that we apply.
Section~\ref{sec:relWork} reviews existing literature related to the dataset reference problem.
All data used by our approach is explained in section~\ref{sec:data}, while section~\ref{sec:approach} introduces the proposed solution.
The evaluation is then presented in section~\ref{sec:eval}.
Section~\ref{sec:future} concludes with an outlook to future work. 
 
\subsection{Problem Statement}
Whereas a lot of effort has been spent on information extraction task in general~\citep{Sarawagi2007}, few attempts have focused on the specific use case of dataset reference extraction (see, e.g., \citep{MeiyuLu2012}). 
When referring to the same dataset, different authors often use different names or keywords.
Although there are proposed standards for dataset citation in full-texts, researchers still ignore or neglect such standards (see, e.g., \cite{altman2007proposed}).
Since the structure of scientific documents does not always follow a standard, and since datasets are rarely being linked to in a consistent way,
simple keyword or name extraction approaches do not solve the problem \citep{Nadeau2007}. 
Table~\ref{table:citation-variety} shows concrete examples of different referencing styles in five articles from the mda journal (see Section~\ref{sec:mda}) that have cited different versions of a study (ALLBUS/GGSS = Allgemeine Bev\"olkerungsumfrage der Sozialwissenschaften/German General Social Survey).

\begin{table}[h!]
	\renewcommand{\arraystretch}{2}
	\centering
	\begin{tabular}{p{.6cm}p{2.5cm}p{7.8cm}}
		\hline
		\bf Article &  \bf Article DOI & \bf Citation style \\
		\hline
		A  & 10.12758/mda.2013.014 &ALLBUS (2010)\\
		
		B  & 10.12758/mda.2013.004 &GESIS -- Leibniz-Institute for the Social Sciences: ALLBUS 2010 -- German General Social Survey. GESIS, Cologne, Germany, ZA4610 Data File version 1.0.0. (2011--05--30), doi:10.4232/1.10445. \\ 
		
		C & 10.12758/mda.2013.012 &ALLBUS (Allgemeinen Bev\"olkerungsumfrage der Sozialwissenschaften)\\
		
		D & 10.12758/mda.2014.007 &(e.g., in the German General Social Survey, ALLBUS; see Wasmer, Scholz, Blohm, Walter and Jutz, 2012)\\
		
		E & 10.12758/mda.2013.019 & Die Einstellungen zu Geschlechterrollen wurden mit Hilfe von Items aus den ALLBUS -- Wellen 1994 und 2008 operationalisiert.\\\hline
	\end{tabular}
	\caption{Citation styles for a study in five different articles.}
	\label{table:citation-variety}
\end{table}

The challenge that our research aims at is to turn each dataset reference detected in an article into an explicit link, for example, using the DOI of the dataset entry in a dataset registry. 
Dataset registries usually provide further metadata about datasets, which can facilitate the detection of such links, such as the dataset creator, publication date, description, and temporal coverage.
In our case, references to datasets should be linked to items in the {\dara} registry, which covers datasets from the social sciences and economics.

\subsection{Contribution}

This article makes the following contributions:
\begin{itemize}
\item a quantitative analysis of typical naming patterns used in the titles of social sciences datasets,
\item a semi-automatic approach for finding references to datasets in social sciences articles with two alternative interactive disambiguation workflows, and
\item an evaluation of the implementation of our approach on a testbed of journal articles.  
\end{itemize}

\section{Preliminaries}
\label{sec:preliminaries}
In this section we introduce key terminology and concepts as well standard metrics for ranking the results of a search query.
Search queries are used for retrieving text fragments in an article that refer to a dataset.

\subsection{Dataset (Terminology)} 
\emph{Dataset} is an ambiguous term; different authors have suggested a great variety of definitions~\cite{peplerpreservation}.
\citeauthor{renear2010definitions} introduced a general notion of the term based on the definitions in the technical and scientific literature.
They define a dataset in terms of its four basic characteristics \emph{grouping}, \emph{content}, \emph{relatedness}, and \emph{purpose}:
\begin{itemize}
	\item \emph{Grouping.} A dataset is considered a \emph{group} of data.
		\emph{Set}, \emph{collection}, \emph{aggregation}, and \emph{atomic unit} are some cases of this feature type.
		For instance, a dataset may have \emph{set} semantics (e.g. “Set of RDF triples”)~\cite{renear2010definitions}, i.e.\ following the mathematical definition of a set, or it may have \emph{collection} semantics, which means that deletion and addition of data does not have any effect on the dataset's identity.
	\item \emph{Content.} The \emph{content} feature describes the data in a dataset.
		For example, \emph{observation} describes propositional content, while \emph{value} refers to measurable content.
	\item \emph{Relatedness.} A dataset is a group of \emph{related} data.
		The \emph{relatedness} feature thus clarifies the inter-relation of data in a dataset.
		\emph{Syntactic} and \emph{semantic} are some examples of this feature.   
	\item \emph{Purpose.} The \emph{purpose} feature refers to the idea of the scientific activity which the dataset is created for.
\end{itemize}

\subsection{Weighting terms in documents using tf-idf}
\label{sec:tfidf}
The \emph{bag of words} model represents a text as a set of terms without considering their order.
Based on this model, documents and queries can be represented in different ways, such as binary vectors, count vectors, and weight vectors. 

In a weight matrix, each row represents a term and each column represents the vector of a document or query. 
Each cell in a weight matrix represents the weight of a term in a document. 
\emph{Tf-idf} (term frequency - inverse document frequency) is one way of computing the weights.

Term frequency (tf) measures the number of occurrences of a given term (t) in a given document (d) or query~\citep{SALTON1988}. 
Tf is calculated using the following formula:
\begin{center} 
	$w_{t,d}$=
	$\begin{cases}
	1+\log_{10} \mathit{tf}_{t,d} & \text{if $\mathit{tf}_{t,d}>0$} \\
	0 & \text{if $\mathit{tf}_{t,d}=0$}\\
	\end{cases}$
\end{center}

The reason for using a logarithm in the formula is that a high number of occurrences of a term does not make the document linearly more relevant.
The total weight for a document with respect to a query is calculated by summing the weights of all query terms in the document.
These terms should appear in both the query $q$ and the document $d$.
Tf is zero if none of the query terms exists in the document.

\begin{center}
	$\mathit{tf\_score}=\sum_{t\in q\cap d} W_{t,d}$
\end{center}

$\mathit{df}_t$ is the number of documents in the corpus that contain $t$, so the more a term is repeated in the corpus, the less informative it becomes.
This leads to the \emph{idf} (inverse document frequency) measure.
Idf is effective for queries that have more than one term.
The following formula defines idf, where $N$ is the number of documents in the corpus.

\begin{center} 
	$\mathit{idf}=\log_{10} (N/\mathit{df}_t)$
\end{center}

\emph{Tf-idf} is defined as the product of \emph{tf} and \emph{idf}. 

\begin{center}
	$\textit{tf-idf}(q,d)=\sum_{t\in q\cap d} \mathit{tf}\cdot\mathit{idf}_{t,d}$
\end{center}

When ranking documents that contain a term being searched, tf-idf returns high scores for documents for which the given term is \emph{characteristic}, i.e.\ documents that have many occurrences of the term, while the term has a low occurrence rate in \emph{all} documents of the corpus.
In other words, tf-idf assigns a weight to each word in a document, giving high weights to keywords and low weights to generally very frequent words.

\subsection{The cosine similarity metric}
\label{sec:cosine}
A Boolean search enables users to find patterns: if a document matches the pattern, the result is “true”; otherwise the result is “false”.
In other words, documents either do or do not satisfy a query expression.
A \emph{ranked} retrieval model is more sophisticated, in that it returns a ranked list of documents in a corpus by considering a query. 

Similarity measures such as Matching, Dice, Overlap Coefficient, and Jaccard are some examples of approaches for ranking a list of documents within a query (cf.~\citet{ChristopherD1999}).
Matching Coefficient finds the numbers of terms that occur in both the query and document vectors.
It calculates the cardinality of the intersection of each document and the query.
\begin{align*}
\mathit{Matching Coefficient}=|d\cap q|
\end{align*}
Dice Coefficient, Overlap Coefficient, and Jaccard try to normalize the Matching Coefficient in different ways:
\begin{itemize}
	\item $\mathit{Dice Coefficient}=\frac{2|d\cap q|}{|d|+|q|}$
	\item $\mathit{Overlap Coefficient}=\frac{|d\cap q|}{\min(|d|,|q|)}$
	\item $\mathit{Jaccard Coefficient}=\frac{|d\cap q|}{|d\cup q|}$
\end{itemize}
Each of them has specific shortcomings;
for example, Jaccard neither applies optimal normalization on the length of documents nor considers term frequency in a document and the corpus of documents.
A document can be converted into a weight vector (see Section~\ref{sec:tfidf}), which looks like $d=(w_1,\dots,w_n)$, and tf-idf is one way of computing the weight $w_i$ of terms.

Search results for a multi-word query in a corpus of documents can be ranked by the similarity of each document with the query.
Given a query vector $q$ and a document vector $d$, their \emph{cosine similarity} is defined as the cosine of the angle $\theta$ between the two vectors \citep{SALTON1988,ChristopherD1999}, i.e.\
\begin{align*}
  \cos(\vec{q},\vec{d})=\cos \theta=\frac{\vec{q}\cdot \vec{d}}{\|\vec{q}\|\,\|\vec{d}\|}=
  \frac{\sum_{i=1}^{|V|} q_id_i}{\sqrt{\sum_{i=1}^{|V|} q_i^2}\sqrt{\sum_{i=1}^{|V|} d_i^2}}
\end{align*}

It normalizes vectors by converting them to their unit vector, thus making documents of different lengths comparable.
Since Euclidean distance is not effective for vectors of different lengths, it ranks documents by angle instead of distance.
Between 0 and 180 degrees, the cosine function decreases monotonically, and therefore larger angles mean less similarity.
Combining tf-idf and cosine similarity yields a ranked list of documents.
In practice, it may furthermore be necessary to define a cut-off threshold in order to distinguish documents that are considered to match the query from those that do not~\citep{Joachims1997}.

\subsection{Precision and recall of a classifier}
\label{sec:precision-recall}
We aim at implementing a binary classifier that tells us whether or not a certain dataset has been referenced by an article.
The algorithm should find references of datasets in an article and then detect a exact match for each detected reference in a text.
These matches are to be selected from titles of datasets in a given dataset registry.

Evaluation metrics such as \emph{precision and recall} determine the reliability of binary classifiers.
For computing a unidimensional ranking of these two dimensions, one typically uses the F-measure, which is defined as the harmonic mean of precision and recall. 
These three metrics are defined as follows~\cite{Powers2011}. 
\begin{itemize}
	\item $\mathit{Precision}=\frac{\#\text{True\ positives}}{\#\text{True positives}+\#\text{False positives}}$
	\item $\mathit{Recall}=\frac{\#\text{True positives}}{\#\text{True positives}+\#\text{False negatives}}$
	\item $\textit{F-measure}=2\cdot{\frac{\mathit{Precision}\cdot\mathit{Recall}}{\mathit{Precision}+\mathit{Recall}}}$
\end{itemize}

If an algorithm returns few wrong predictions, it will lead to a high precision.
An algorithm should predict most of the relevant results to achieve a high recall.

\section{Related work}
\label{sec:relWork}
While only a few works address the specific task of extracting dataset references from scientific publications, a lot of research has been done on its general foundations including metadata extraction and string similarity algorithms. 
Related work can be divided into three main groups covered by the following subsections.
\subsection{Methods based on the “bag of words” model}
One group of methods is based on the “bag of words” model using algorithms such as tf-idf to adjust weights for terms in a representation of texts as vectors (cf.\ the introduction in Section~\ref{sec:tfidf}).
\citeauthor{Lee2008} proposed an unsupervised keyword extraction method by using a tf-idf model with some heuristics~\cite{Lee2008}.
Our approach uses similarity measures for finding a perfect match for each dataset reference in an article by comparing titles of datasets in a registry to sentences in articles.
Dice, Jaccard and Cosine can be applied to a vector representation of a text easily (cf.~\citet{ChristopherD1999}).
The accuracy of algorithms based on such similarity measures can be improved by making them semantics-aware, e.g., representing a set of synonyms as a single vector space dimension.

\subsection{Corpus and Web based methods}
Corpus and Web based methods often use information about the co-occurrence of two texts in documents, and are used for measuring texts' semantic similarity.
\citeauthor{Turney2001} introduced a simple unsupervised learning algorithm for detecting synonyms~\cite{Turney2001}, which searches queries through an online search engine and analyses the results.
The quality of the algorithm depends on the number of search results returned.  

\citeauthor{sighal2013} proposed an approach to extract dataset names from articles~\cite{sighal2013}.
They employed the normalized Google distance algorithm (NGD), which estimates both the probability of two terms existing separately in a document, as well as of their co-occurrence. 
\begin{align*}
	\mathit{NGD}(x,y)=\frac{\max(\log f(x),f(y))-\log f(x,y)}{\log M -\min(\log f(x),\log f(y))}
\end{align*}
In this formula, $M$ is the number of all web pages searched.
$F(x)$ means the number of returned pages for $x$ as a query term and $f(x,y)$ represents the number of pages for the intersection of $x$ and $y$. 
They used two academic search engines -- Google Scholar and Microsoft Academic Search -- instead of a local corpus.

\citeauthor{Schaefer2014} proposed the Normalized Relevance Distance (NRD)~\cite{Schaefer2014}.
This metric measures the semantic relatedness of terms.
NRD is based on the co-occurrence of terms in documents, and extends NGD by using relevance weights of terms.
The quality of these methods depends on the size of the corpus used.

\citeauthor{Sahami2006} suggest a similarity function based on query expansion~\cite{Sahami2006}.
Their algorithm determines the degree of semantic similarity between two phrases.
Each of these phrases is searched by an online search engine and then expanded by using returned documents.
Afterwards, the new phrases are used for computing similarity. 

The problem that we aim to solve involves the two subtasks of 1.\ identifying dataset references in an article, and then 2.\ finding at least one correct match for each of these identified references.
Literature citation mining is the process of determining the number of citations that a specific article receives.
It constructs a literature citation network, which can be used for detecting the impact of an article~
\cite{Afzal2010}.
Citation mining can usually be handled by three subtasks.
First, literature references should be extracted from the bibliography section of a document, and afterwards, metadata extraction should be applied on the references extracted in the first phase.
Finally, each reference should be linked to the cited article by using the metadata extracted in the second step~\cite{Afzal2010}.

Dataset and literature citation mining from documents cannot generally be compared in the detecting phase, since dataset mining needs to be applied to the entire article, but literature mining mostly focuses on the bibliography of the article. 
Unlike the detection phase, they can mostly use the same strategy for the matching phase. 

Afzal et al.\ proposed a rule-based citation mining technique~\cite{Afzal2010}.
Their approach detects literature references from each document and then extracts citation metadata from each of them, such as title, authors, and venue. Based on the venue, it then extracts all related titles from the DBLP computer science bibliography, which contains more than three million articles.
Finally, it tries to link the title of each extracted literature reference and titles found in DBLP.

\subsection{Machine learning methods}
Many different machine learning approaches have been employed for extracting metadata, and in a few cases also for detecting dataset references.
For example, \citeauthor{Zhang2006} \cite{Zhang2006} and \citeauthor{Han2003} \cite{Han2003} proposed keyword extraction methods based on support vector machines (SVM). 

\citeauthor{Kaur2010} conducted a survey on several effective keyword extraction techniques, such as selection based on informative features, position weight, and conditional random field (CRF) algorithms~\cite{Kaur2010}.
Extracting keywords from an article can be considered as a labeling task.
CRF classifiers can assign labels to sequences of input, and, for instance, define which parts in an article can be assumed to be keywords~\citep{ZHANG2008}.

\citeauthor{Cui2010} proposed an approach using Hidden Markov Models (HMM) to extract metadata from texts~\cite{Cui2010}. 
HMM is a language-independent and trainable algorithm~\cite{Kubala1998}.
\citeauthor{Marinai2009} described a method for extracting metadata from documents by using a neural classifier~\cite{Marinai2009}.
Kern et al.\ proposed an algorithm that uses a maximum entropy classifier for extracting metadata from scientific articles~
\cite{Kern2012}.
\citeauthor{MeiyuLu2012} used the feature-based Llama classifier for detecting dataset references in documents~\cite{MeiyuLu2012}.
Since there are many different styles for referencing datasets, large training sets are necessary for these approaches.

\citeauthor{Boland2012} proposed a pattern induction method for extracting dataset references from documents in order to overcome the necessity of such a large training set~\cite{Boland2012}.
Their algorithm starts with either the name of a dataset or with an abbreviation of this name, and then drives patterns of all phrases that contain that name or abbreviation in articles.
The patterns are applied to articles in order to extract more dataset names and abbreviations.
This process repeats with new abbreviations and names until no more datasets can be detected in articles.
It derives patterns of phrases that contain dataset references iteratively by using a bootstrapping approach.

\section{Data sources}
\label{sec:data}
This section describes the three types of data sources that we use. 
We use full-text articles from the \emph{mda journal} to evaluate the performance of our dataset linking approach, and metadata of datasets in the {\dara} dataset registry to identify datasets. 
Finally, we use metadata of the articles registered in the \emph{SSOAR repository}\footnote{\url{http://www.ssoar.info}} for exporting the dataset links suggested for an article in the JSON exchange format.
 
\subsection{Articles from mda journal}\label{sec:mda}
 
Methods, data, analyses (mda\footnote{\url{http://www.gesis.org/en/services/publications/mda/}}) is an open access journal that focuses on research questions related to quantitative methods, with a special emphasis on survey methodology.
It has published research on all aspects of science of surveys, be it on data collection, measurement, or data analysis and statistics.
For our research we used a random sample of full-text articles from mda as test corpus.
 
\subsection{The {\dara} dataset registry}
  
\subsubsection{Overview}
Our approach focuses on social sciences datasets since it uses registered datasets in {\dara} registry\footnote{\url{http://www.da-ra.de}}.
The registry offers a DOI registration service for datasets in the social sciences and economics. 
Different institutions have collected research data in the social sciences and made them available.
Although the accessibility of such datasets for further analyses and other reuse is important, information on where to find and how to access them is often missing in articles.
{\dara} makes 
data referable 
by assigning a DOI to each dataset.
In November 2016, {\dara} holds 486,233 records such as datasets, texts, collections, videos, and interactive resources, more than 35,000 of which are datasets. 
For each dataset, {\dara} provides metadata including title, author, language, and publisher.
This metadata is exposed to harvesters employing a freely accessible API using OAI-PMH (Open Archives Initiative Protocol for Metadata Harvesting\footnote{\url{http://da-ra.de/oaip/}}). 
  
\subsubsection{Analysis of dataset titles in {\dara}}
We analyzed the titles of all datasets in {\dara} after harvesting using the {\dara} API.
The analysis shows that about one third of the titles follow a special pattern, which makes them easier to be detected in the text of an article.
We have identified three such special patterns:
\begin{enumerate}
	\item \emph{Abbreviations} contained in titles, which are typically used to refer to the datasets. 
		Consider, for example, the full title \enquote{Programme for the International Assessment of Adult Competencies (PIAAC), Cyprus}, which contains the abbreviation \enquote{PIAAC}.
	\item \emph{Filenames} as in \enquote{Southern Education and Racial Discrimination, 1880--1910: Virginia: VIRGPT2.DAT}, where \enquote{VIRGPT2.DAT} is the name of the dataset file.
	\item \emph{Phrases} that explicitly denote the existence of dataset references in a text, such as \enquote{Exit Poll} or \enquote{Probation Survey}.
\enquote{Czech Exit Poll 1996} is an example of such a dataset title. 
\end{enumerate}

Abbreviations and special phrases can be found in about 17\% and 19\% of the {\dara} dataset titles respectively.
The intersection of these two groups only accounts for 1.49\%.
Filenames occur in less than one percent of the titles. 
The proposed approach in this article uses only the first and the last category, since the filename category only covers a small amount of titles.

\subsection{The Social Science Open Access Repository (SSOAR)}
The SSOAR repository provides full-text access to documents.
It covers scholarly contributions related to different social science fields such as social psychology, communication sciences, and historical social research.
Today, approx.\ 38,000 full-texts are available.
This repository provides some metadata such as abstract and keywords for each article in both German and English.
It is assumed as a secondary publisher which publishes pre-prints, post-prints, and original publishers' versions of scholarly articles but it also lets authors publish their work for the first time.
Furthermore, the metadata of the articles inside the repository can be harvested.
SSOAR assigns a URN (Uniform Resource Name) as a persistent identifier (PID) to each full-text to establish a stable link to the article.
If the full-text is the pre-print or post-print version of a published work, the repository uses the DOI of the article.

\section{A semi-automatic approach for finding dataset references}
\label{sec:approach}
We have realized a semi-automatic approach for finding references to datasets registered in {\dara} in a given full-text. 
Our approach is divided into four main steps.
The first step is related to generating special features dictionaries from datasets' titles.
The second step deals with identifying and matching references to datasets in an article, and the third step focuses on improving the results of the second step.
In the final step a user can export the results.

It took a semi-automatic approach since 
the first and last steps of our algorithm require human interaction to improve the accuracy of the result. 
In the first step, the user should review two generated lists of abbreviations and special phrases.
In the final step, the user should make the final decision regarding the references suggested by our approach.

The main differences between our approach and related ones are that ours neither requires a huge corpus of articles nor a large training set.
Our approach is straightforward and able to prepare results for a given article in a few minutes or even seconds depending on the number of datasets' references in the article that we want to analyze.

\subsection{Step 1: Preparing the dictionary}
\label{sec:preparing-dictionary}
The preparation of a \emph{dictionary} of abbreviations and special phrases is the first step.
\emph{Abbreviations} are initially obtained by applying certain algorithms and rules to the dataset titles harvested from {\dara}.
The dataset titles are preprocessed automatically before the abbreviations are extracted.
Titles fully in capital letters are removed, the remaining titles are split based on \enquote{:}, and of titles that contained a colon only the first parts are kept.
The extraction of abbreviations from dataset titles follows specific steps:

\begin{enumerate}
	\item The titles are tokenized (by using nltk -- a Python package for natural language processing).
	\item The tokens that are not completely in lowercase (not including the first letter) -- not only a combination of digits and punctuation marks, not Roman numerals, and do not start with a digit are added to a new list (e.g. \enquote{SFB580-B2}, \enquote{A*CENSUS}, \enquote{L.A.FANS}, \enquote{aDvANCE}
	and \enquote{GBF/DIME}).
	\item The titles are split based on `-' and `(', and then single tokens before such delimiters are added to the list (e.g. \enquote{euandi} in \enquote{euandi (Experteninterviews) - Reduzierte Version}).
	\item The items on the list of abbreviations should only contain the punctuation marks `.', `-', `/', `*' and `\&'. (e.g. \enquote{NHM\&E}).
	\item The items that contain `/' or `-' and are also partially in lowercase are removed from the list (first letter of each part is not included; e.g.\ \enquote{Allbus/GGSS} is removed). 
	\item Words in German and English, as well as country names, are removed from the list. Words, fully or partially in capital letters will not be pruned by dictionary (first letter is not included).
\end{enumerate}
The titles fully in capital letters are converted to lowercase and tokenized.
Afterwards, their tokens without definition are added to the list.
These algorithms and rules correctly detect, for example, \enquote{DAWN} in \enquote{Drug Abuse Warning Network (DAWN), 2008}.
However, it sometimes detects abbreviations that are not references to datasets, such as \enquote{NYPD} in \enquote{New York Police Department (NYPD) Stop, Question, and Frisk Database, 2006}.
As their identification is hard to automate, we assigned this task to a human expert. 
The expert reviews the list and then makes a false positive list -- afterward, the list will be removed from the dictionary automatically.
In an experiment involving all dataset titles of {\dara}, we notice the ratio of false positives is one out of three items of the list of abbreviations extracted automatically.
This means that approximately two thirds of the abbreviations are derived correctly, and pruning the rest of the titles from the list requires little effort.

The preparation of the dictionary of special phrases also needs human interaction.
A list of terms that refer to datasets such as \enquote{Study} or \enquote{Survey} has been generated manually; this list contains about 30 items.
Afterwards, phrases containing these terms were derived by some algorithms and rules from the titles of actual datasets in {\dara}.
Three types of phrases are considered here, the first of which are tokens that include a dictionary item such as \enquote{Singularisierungsstudie}, which contains \enquote{studie}.
The second is a category of phrases that includes \enquote{Survey of} or \enquote{Study of} as a sub phrase as well as one more token that is not a stop word, such as \enquote{Survey of Hunting}.
The last one is phrases that contain two tokens, where one of them is a dictionary item such as \enquote{Poll}, and the second token should not be a stop word such as \enquote{Freedom Poll}. 

A human expert has finally verified the phrase list, and false positives are added to the list which contains false positive items.
In the phrase list, there are few false positives, and most can be detected by a human expert 
while our algorithm processes articles.
Users should decide about the output so each time a false result generated by a false phrase occurs, the users can easily add the phrase to the false positive list. This means our approach will improve over time.
 
\subsection{Step 2: Detecting dataset references and ranking matching datasets}
\label{sec:detecting-ranking}
Next, the characteristic features (abbreviations or phrases) of dataset titles are detected in the full-text of a given article.
The text is split into sentences, and each of these features is searched for in each sentence. 
Any detection of the special features in a text means a dataset reference in the text exists.
A sentence is split into smaller pieces if a feature repeats inside the sentence more than once, since such a sentence may contain references to
different versions of a dataset.
Any phrase identified in this step might correspond to more than one dataset title.

For example, \enquote{ALLBUS} is an abbreviation for a famous social science dataset, of which more than 150 versions are registered in {\dara}.
These versions have different titles and, for instance, the titles differ from year of study such as \enquote{German General
Social Survey -- ALLBUS 1998}, 
\enquote{German General Social Survey -- ALLBUS 2010}, and \enquote{German General Social Survey (ALLBUS) -- Cumulation 1980--2012}.
In another example, two titles that both contain the \enquote{PIAAC} abbreviation are \enquote{Programme for the International Assessment of Adult Competencies (PIAAC), Cyprus} and \enquote{Programme for the International Assessment of Adult Competencies (PIAAC), Germany}, i.e., two datasets that differ in their geographic coverage.
The last example is given by two versions of the \enquote{EVS} dataset, 
\enquote{EVS -- European Values Study 1999 -- Italy} and \enquote{European Values Study 2008: Azerbaijan (EVS 2008) }, which differ in both their year of study and geographic coverage.

We solve the problem of identifying the most likely datasets referenced in the article by ranking their titles with a combination of tf-idf and cosine similarity.
In this ranking algorithm, we apply the definitions of Section~\ref{sec:preliminaries}, where the query is a candidate dataset reference found in the article and the documents are the titles of all datasets in {\dara}. 
It means that our approach tries to identify the most similar dataset title in the {\dara} repository with a sentence that contains any of the special features where the sentence belongs to the analyzed article.

\subsection{Step 3: Heuristics to Improve Ranking in Step 2}
\label{sec:heur-impr-rank}

For each reference detected in the full-text of an article we compute tf-idf over the full-text and the list of dataset titles in {\dara}, which contain a specific characteristic feature (abbreviation or phrase) detected in the reference.

As we observed that it leads to many false positives, comparing all datasets' titles with a sentence in an article, and, afterwards, ranking titles based on their score was not useful.
We solved these problems by involving special features.

Our approach considers only the list of titles that contain the special feature detected in the reference, since they are related titles and the rest of the titles in the registry is irrelevant.
We limit our options in order to improve the accuracy of our approach.
We decided to use the list of titles and whole sentences of the article, and not only the reference sentence, since this consideration enables us to have a bigger corpus of documents and to obtain a better weight for each word.
The utilization of titles that contain special features reduces the weight score of the feature and raises the weight scores of other terms in the reference sentence.
It therefore has a positive impact on accuracy.

\begin{figure}[h]
	\centering
	\includegraphics[width=4.5 in]{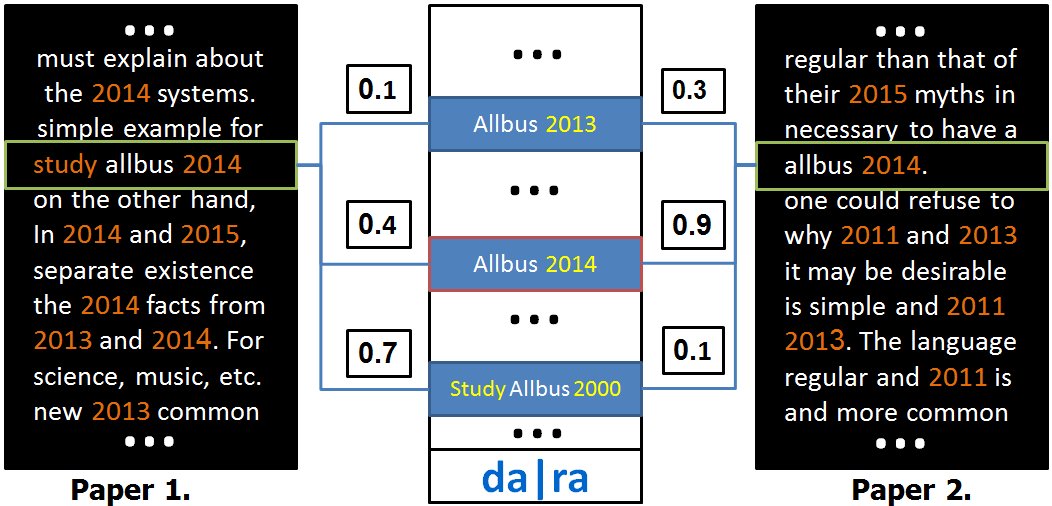} 
	\caption{A toy example of cosine similarity, where tf-idf is computed over phrases in two articles.}
	\label{fig:similarity-example}
\end{figure}

While a corpus of articles is typically huge, the size of all {\dara} dataset titles and the size of the full-text of an average article are less than 4 MB each.
Given this limited corpus size, our algorithm may detect some false keywords in a query, thus adversely affecting the result.
Figure~\ref{fig:similarity-example} illustrates a toy example of this problem.

In paper~1, \enquote{2014} repeats many times, whereas the word \enquote{study} occurs only once, which means the tf-idf assigns a high weight to \enquote{study} and a low weight to \enquote{2014}. When the query string is \enquote{study allbus 2014}, cosine similarity gives a higher rank to \enquote{Study Allbus 2000} than \enquote{Allbus 2014}.
To address this problem in a better way, our implementation employs some heuristics.
These include an algorithm that improves dataset rankings based on matching years in the candidate strings in both the article and the datasets' titles.
In the example, these heuristics improve the ranking of the \enquote{Allbus 2014} dataset when analyzing paper~1. 

\begin{figure}[h]
	\centering
	\includegraphics[height=3in]{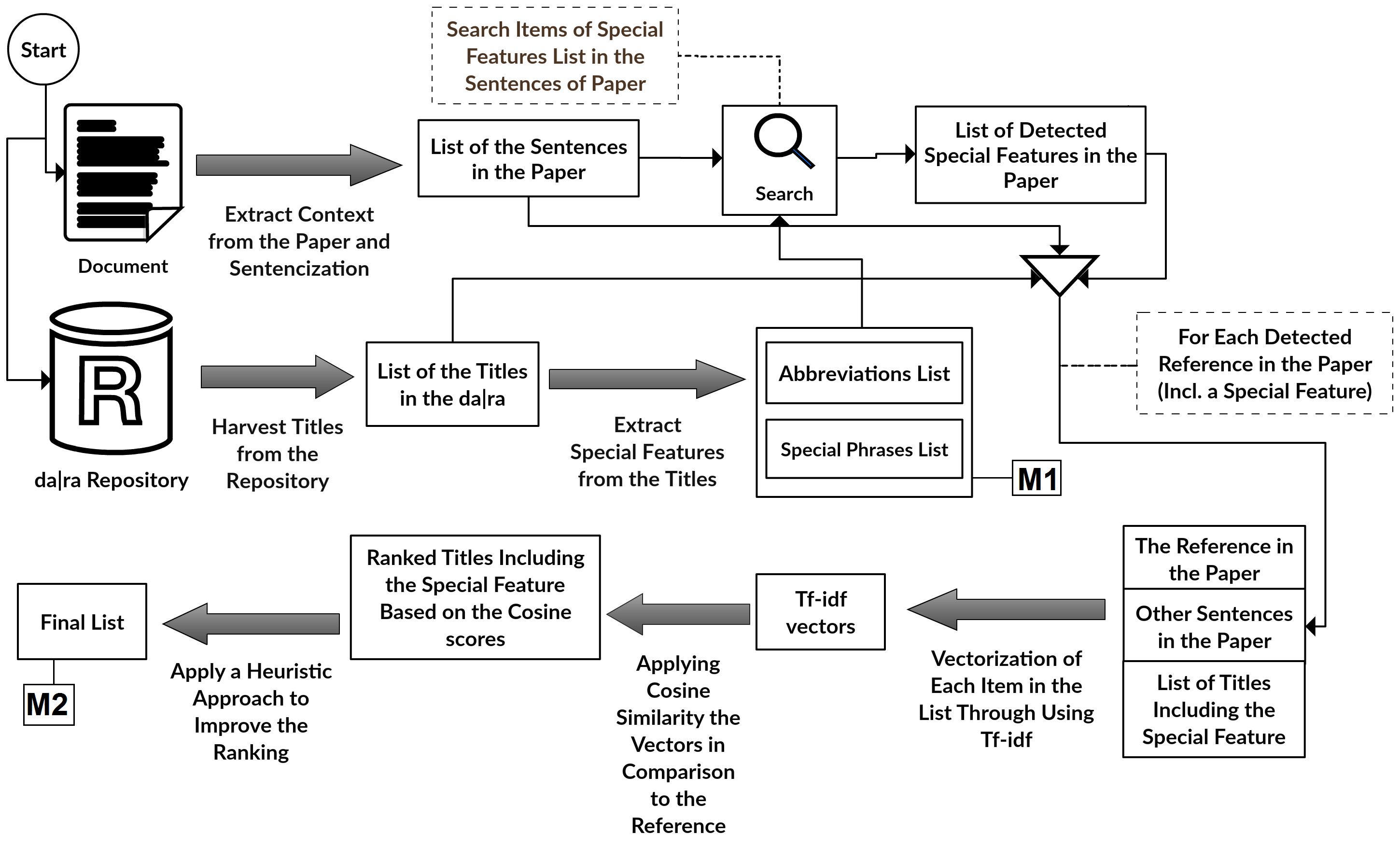}
	\caption{An overview of the approach.}
	\label{fig:Overview_of_approach}
\end{figure}

Figure~\ref{fig:Overview_of_approach} shows an overview of our approach.
The two steps labeled with \enquote{M} require human interaction.
\enquote{M1} is about the preparation of lists of special features and \enquote{M2} is about making final decisions between candidates suggested by our approach.

\subsection{Step 4: Exposing the Results to the User, and Interactive Disambiguation}
\label{sec:expos-results-reus}	
The application of our approach supports two workflows through which an expert user can choose the best matches for the datasets cited by an article from a set of candidates identified automatically. The sizes of these sets have been chosen according to the observations we made during the evaluation of the automated step, as explained in Section~\ref{sec:eval}.

One workflow works per reference: for each reference, five titles of candidate datasets are suggested to the user.
While this workflow best supports the user in getting every reference right, it can be time consuming.
Each article in our corpus contains 45 dataset references on average, but these \emph{references} only belong to an average number of 4 distinct \emph{datasets}.

The second alternative workflow takes advantage of this observation.
It works per characteristic feature and suggests 6 titles of candidate datasets to the user for each feature (which may be common to multiple individual references in the article). 

Finally, an RDF graph containing information about links identified between articles and datasets is exported as an output.
To enable even further analysis of the links identified between articles and datasets, we export an RDF graph containing all candidate datasets identified in the latter workflow for each article.
For each candidate dataset, we represent the essential metadata of the dataset in RDF: DOI and title.

\section{Evaluation}
\label{sec:eval}
 \label{sec:eval}
 The computation of evaluation metrics such as precision, recall, and F-measure require a ground truth.
 We therefore selected a test corpus of 15 random articles\footnote{\url{https://github.com/OSCOSS/testdata_mda}} from the 2013 and 2014 issues of the mda journal (see Section~\ref{sec:mda}) -- 6 in English and 9 in German. 
This test corpus includes 25 datasets without considering different versions of each dataset (see more information about the test corpus in Table~\ref{table:info-testcorpuse}).

A trained assessor from the InFoliS\footnote{Integration von Forschungsdaten und Literatur in den Sozialwissenschaften = integration of research data and literature in the social sciences} II project at GESIS reviewed all articles one by one and identified all references to datasets (see second row in table~\ref{table:info-testcorpuse}).
Afterwards, the assessor attempted to discover at least one correct match in {\dara} for each detected reference, resulting in a list of correct datasets per article.
These lists were used as a gold standard\footnote{\url{https://github.com/OSCOSS/testdata_mda/blob/master/ELPub_Corpus_evaluation.csv}} to compare with the results of our algorithm to examine differences and similarities.

\begin{table}[h!]
	\renewcommand{\arraystretch}{2}
	\centering
	\begin{tabular}{c c c c}
		\FL
		 & Max. & Min. & Avg.
		\ML
		\# of datasets in an article & 7 & 1 & 4
		\NN
		\# of references to datasets & 147 & 1 & 45
		\LL
	\end{tabular}
	\caption{Test corpus}
	\label{table:info-testcorpuse}
\end{table}

\subsection{Evaluation Process and Description}
We decided to divide our evaluation into two steps.
The first step focuses on identifying dataset references in articles.
Here, accuracy depends on the quality of the generated dictionaries of abbreviations and special phrases 
(the accuracy metrics used in the article are explained in Section~\ref{sec:precision-recall}).
 
Our algorithm searches these characteristic features (as explained in Section~\ref{sec:detecting-ranking}) in the full-texts; detection of any of these features may lead to the detection of a dataset reference (see row \enquote{Detection} in Table~\ref{table:eval-results}).
In this phase, if a characteristic feature is identified both in an article and in the gold standard, it will be labeled as a true positive.
If the feature is in the gold standard but not in our output, it will be labeled as a false negative, or as a false positive in the opposite case.
 
The second step of the evaluation is about the accuracy of matching detected references in articles with datasets titles in the {\dara} registry.
This evaluation phase considers only true positives from the previous step.
The lists of suggested matches for an item, both from the gold standard and from our output, are compared in this step. 
Since a dataset may occur on its own or be integrated together with other studies,
an item can have more than one true match (e.g. Allbus 2010 in ALLBUScompact 1980--2012).
In this step, an item will be labeled as a false negative if none of the suggestions for the item in the gold standard appears in the output of our algorithm.
The numbers of false positives and false negatives are equal in this step, since a missing corresponding match means the existence of false positives.
True positives, false positives, and false negatives are counted and then used to compute precision and recall.

The third row in Table~\ref{table:eval-results} refers to the accuracy of two phases of the algorithm as one unit in order to find how well it works generally, and does not consider one specific phase (i.e. identification or matching).
In order to satisfy this purpose, we repeated the second phase of the evaluation, but this time included all data from the first step and not only the true positives.
If an item is identified as false positive in the first section of evaluation, it is labeled as such in the evaluation as well.   
 
\subsection{Evaluation Results}
\label{sec:evre}
The algorithm gains high precision in both the detection and matching phases, which means it has a small number of wrong predictions.
It also covers the majority of relevant datasets, which leads to high recall.
The results of our evaluations are shown in Table~\ref{table:eval-results}.
 
\begin{table}[h!]
 	\renewcommand{\arraystretch}{2}
 	\centering
 	\begin{tabular}{c c c c}
 		\FL
 		Phase of Evaluation & Precision & Recall & F-measure
 		\ML
 		Detection & 0.91 & 0.77 & 0.84
 		\NN
 		Matching & 0.83 & 0.83 & 0.83
 		\NN
 		Detection+Matching & 0.76 & 0.64 & 0.7
 		\LL
 	\end{tabular}
 	\caption{Results of the evaluation}
 	\label{table:eval-results}
\end{table}
 
Our observations in the second evaluation step confirm the choices of set size in the interactive disambiguation workflows.
In the per-reference matching workflow (as mentioned in Section~\ref{sec:expos-results-reus}), a ranked list of dataset titles is generated for each of the 45 dataset references (on average in our corpus) in an article by employing a combination of cosine similarity and tf-idf. 
 
Our observation shows that the correct match among {\dara} dataset titles for each reference detected is in the top 5 items of the ranked list generated by combining cosine similarity and tf-idf for that reference.
Therefore, we adjusted our implementation to only keep the top 5 items of each candidate list for further analysis, such as an expert user's interactive selection of \emph{the} right dataset for a reference.
 
The per-feature matching workflow (as mentioned in Section~\ref{sec:expos-results-reus}) categorizes references by characteristic features.
For example, in an article that contains exactly three detected characteristic features -- \enquote{ALLBUS}, \enquote{PIAAC}, and \enquote{exit poll} -- each dataset reference relates to one of these three features.
If we obtain for each such reference the list of top 5 matches as in the per-reference workflow and group these lists per category, we can count the number of occurrences of each dataset title per category. 
 
Looking at the dataset titles per category sorted by ascending number of occurrences, we observed that the correct matches for the datasets' references using a specific characteristic feature were always among the top 6 items.
 
\section{Conclusion and future work}
\label{sec:future}
We have presented an approach for detecting references to datasets in social sciences articles.
The approach works in real time and does not require any training dataset.
There are just some manual tasks such as initially cleaning the dictionary of abbreviations, or making final decisions among multiple candidates suggested for the datasets cited by the given article.
In our evaluation we have achieved an F-measure of 0.84 for the detection task and an F-measure of 0.83 for finding correct matches for each reference in our gold standard.  
Although the {\dara} registry is large and it is growing, there are still many datasets that have not yet been registered there. 
This circumstance will adversely affect the task of detecting references to datasets in articles and matching them to items in {\dara}.
After the evaluation, our observations reveal that {\dara} could cover only 64\% of datasets in our test corpus. 

Future work will focus on improving the accuracy of detecting references to the datasets supported so far, and on extending the coverage to all datasets.
Accuracy can be improved by better similarity metrics, e.g., taking into account synonyms and further metadata of datasets in addition to the title.

Coverage can be improved by taking into account further datasets, which are not registered in {\dara}.
One promising further source of datasets is OpenAIRE, the Open Access Infrastructure for Research in Europe, which so far covers more than 16,000 datasets from all domains inluding social science but is rapidly growing thanks to the increasing attention paid to open access publishing in the EU.
The OpenAIRE metadata can be consumed via OAI-PMH, or, in an even more straightforward way, as linked data (cf.\ our previous work, \citet{VahdatiEtAl:MappingResearchMetadata15}).

For reuse in subsequent analytics we are planning to enrich the RDF export.
Currently, it neither includes the exact positions of dataset references in an article, nor our algorithms' confidence in every possible matching dataset.
While we are already obtaining in the course of the per-reference matching workflow (cf.\ Section~\ref{sec:expos-results-reus}), we will additionally preserve it in the per-feature workflow, from which the RDF export is generated.

In a mid-term perspective, solutions for identifying dataset references in articles that have been published already could be made redundant by a wider adoption of standards for properly citing datasets while authoring articles, and corresponding tool support for authors.

\paragraph{Acknowledgments}
This work has been partly funded by the DFG project “Opening Scholarly Communication in Social Sciences” 
(grant agreements SU 647/19-1 and AU 340/9-1), and by the European Commission under grant agreement 643410. 
We thank Katarina Boland from the InFoLiS II project (MA 5334/1-2) for helpful discussions and for generating the gold standard for our evaluation. 
This article is an extension of our previous work~\cite{ghavimi2016identifying}.
The implementation code is available in the OSCOSS repository (link: \url{github.com/OSCOSS/Dataset_Detcter/tree/master/src}). 

\bibliography{biblo}
\end{document}